# Emergence of integrated institutions in a large population of self-governing communities

Seth Frey*,[1,2], Robert W Sumner[3]

[1] Communication Department, University of California Davis, Davis, CA, USA

[2] Neukom Institute for Computational Science, Dartmouth College, Hanover, NH, USA

[3] Computer Science, ETH Zurich, Zürich, Switzerland

* Corresponding author

E-mail: sfrey@ucdavis.edu (SF)

# Abstract

Most aspects of our lives are governed by large, highly developed institutions that integrate several governance tasks under one authority structure. But theorists differ as to the mechanisms that drive the development of such concentrated governance systems from rudimentary beginnings. Is the emergence of integrated governance schemes a symptom of consolidation of authority by small status groups? Or does integration occur because a complex institution has more potential responses to a complex environment? Here we examine the emergence of complex governance regimes in 5,000 sovereign, resource-constrained, self-governing online communities, ranging in scale from one to thousands of users. Each community begins with no community members and no governance infrastructure. As communities grow, they are subject to selection pressures that keep better managed servers better populated. We identify predictors of community success and test the hypothesis that governance complexity can enhance community fitness. We find that what predicts success depends on size: changes in complexity predict increased success with larger population servers. Specifically, governance rules in a large successful community are more numerous and broader in scope. They also tend to rely more on rules that concentrate power in administrators, and on rules that manage bad behavior and limited server resources. Overall, this work is consistent with theories that formal integrated governance systems emerge to organize collective responses to interdependent resource management problems, especially as factors such as population size exacerbate those problems.

# Introduction

The Internet has empowered normal people to easily craft and deploy whole social systems, and to attract and maintain vibrant communities of total strangers. As online communities thrust millions around the world into the shared management of such artificial "resource systems", these unwitting amateur governors find themselves struggling to manage a disparate bundle of social dilemmas. Wikipedia, for example, must reduce vandalism from malicious edits, retain quality editors, and solve the public goods problem of soliciting contributions to a service that is free to all. Studying communities' failures and successes promises not only to advance fundamental questions about resource governance [1,2], but to cultivate in citizens a native comfort with the skills, in both governing and being governed, that demanding institutional forms like democracy require [3].

Understanding success in online communities is in large part understanding the development of stable formal governance systems from rudimentary beginnings. What drives institutions to expand from individual collective action problems and consolidated themselves into bodies that integrate operations over complexes of governance tasks? Some political and organizational theories characterize the expansion of institutions beyond simple idealized forms as a symptom of organizational pathologies, as special interests capture governance mechanisms, or historical accidents accumulate inefficiently [4-6]. These theories, typical of anarchist and libertarian thought, see formal governance as emerging to institutionalize the power of small status groups, or more efficiently extract rents. A less cynical view, the theory of "institutional diversity," holds that governance institutions grow in size and complexity in order to expand their purview and more effectively respond to interdependent environmental demands [7,8]. This latter idea, an application of cyberneticist W. Ross Ashby's Law of Requisite Variety [9,10], captures an observation from decades of fieldwork in forests, fisheries, pastures, and irrigation systems, that the most successful and enduring self-governing resource management institutions are those that use a variety of rule types, that integrate a range of collective tasks under one institutional structure, and that distribute authority across many levels of that structure [7,11]. The thesis of this perspective is that the resource management systems that communities develop are as complex as they need to be to adapt to their socio-ecological circumstances [12].

Insights into the evolution of governance complexity are necessary for understanding effective policies and organizations. And questions around the origin and success of complex, integrated, or formal institutions are of immense theoretical interest across the social sciences, in the institutional, historical, and organizational subdisciplines of economics [13-15], anthropology [16], political theory [17,18], organization theory [19], from animal behavior [20] to complex systems [21-25]. In pursuit of these theories, empirical researchers have benefitted from simple artificial norm systems that focus on a single problem [26,27] and, at the other extreme, from individual quantitative case studies of mature multifaceted institutions [28-30]. But closing the explanatory gap between these extremes — explaining the development of fully developed institutions from rudimentary emergent norm systems — is a forbidding challenge, one that demands population-level comparisons across many independent but comparable institutions [1,31-35]. Independent hosts of Wikipedia's MediaWiki software platform have been an especially fruitful domain for comparative system-scale analysis, with large scale comparisons illuminating questions around structural predictors of successful online communities, the emergence of oligarchy in online peer production systems, and their life histories [36-38]. Virtual social systems like peer production platforms, forum communities, and multiplayer game worlds are appealing because they are relatively simple, highly replicable, easily scrapeable, and they attract motivated people pursuing clear goals under knowable constraints. Our analysis of the rule systems of 5,216 virtual communities within one popular online platform contributes to an understanding of comparative institutional design by revealing correlates of community building success, and suggesting mechanisms for those relationships in terms of the theory of institutional diversity. Our approach, an automated population-level comparative analysis of standardized sociotechnical systems, is naturalistic and yet constrained enough by theory and in form to be amenable to causal interpretation.

# The resource management perspective

The frameworks of natural resource management are valuable for understanding successful administration of online communities. We focus on the frameworks of the Ostrom Workshop, which have been used to study other kinds of online communities [39-43].

Community building, online and in general, is attended by a number of collective action problems that derive from the special properties of different resources, particularly the properties of non-excludability and subtractability. A non-excludable resource is one that is not subject to private ownership in the sense that an agent cannot prevent others from accessing or consuming that resource. A subtractable resource is one that is finite: consumption by one agent constrains the potential consumption of it by another. Common pool resources are defined as non-excludable and subtractable [8], and examples include common pastures, shared office kitchen sinks, and global fish populations. They are vulnerable to over-extraction via the mechanism of the Tragedy of the Commons [44], a property that unifies otherwise disparate commons such as fisheries, forests, and irrigation systems [32]. Public goods are also non-excludable but they differ from common pool resources in being defined by their non-subtractability. Examples include breathable air in a room, radio broadcasts, uncongested public roads, and not-for-profit online information sources. Because they are not limited, the social dilemma commonly associated with public goods is less concerned with regulating access than with ensuring sufficient provisioning. Public goods often require maintenance, but the same incentives that drive individuals to over-extract common pool resources drive them to neglect public goods [45]. Public bads are a type of resource with the same properties as public goods except that provisioning tends to be costless and their externalities are negative rather than positive; public bads are pollutants, and the collective action problem associated with them is pollution, costless to produce, costly to manage, and affecting everyone. An effective institution will solve public bad provisioning by minimizing its costs or discouraging the behaviors that produce it.

Understanding a community's resources in these abstract terms makes community success amenable to frameworks for analyzing real world community resource management institutions such as local fisheries, irrigation systems, and forest management communities. These systems differ dramatically in their social and ecological particulars, but researchers have succeeded at articulating theories and approaches that make lessons from one generalizable to others. It is from these generalization efforts that we have a basis to argue for the generality of our conclusions: as online server communities can be treated as resource management institutions that manage familiar resource types, and mitigate familiar problems, the factors that lead to their success are promising candidates as predictors of success in other resource settings.

## The challenge of governing a sovereign online community

The millions of sovereign self-governing communities on the Web provide an unprecedented opportunity for comparative insights into institutional development over multiple scales of population [35,46,47]. Any person with an Internet connection can follow simple templates to build a global community around common interests. And just as in a physical community, success online depends on a community's ability to overcome collective action problems of longstanding scientific interest [11,44,48]. In this sense, online communities are self-organized resource governance systems that are replicable, and that can advance existing theories of institutional success [32].

Over a period of two years, we monitored over 5,000 web servers hosting instantiations of a popular online community platform (described in detail in the Methods and SI Text), in order to compare the success of their communities. Server communities in this system are all sovereign and independent from each other, yet they face common resource management problems that put them under intense selection for survival. They face daunting resource constraints that are amplified by population growth. They share standardized interfaces with the outside world, yet are customizable enough to permit vast individual differences in culture and management style. And they are all led by their founding server administrator, who by default makes all governance choices unilaterally and bears all time and money costs. In this setting, success means recruiting, supporting, and retaining a core group of devoted community members and fostering successful collective action among them. In order to attain it, an administrator must overcome the resource challenges that cause communities to fail.

After an administrator specifies the desired size of their community (ranging in our data from the single digits to the hundreds), they are responsible for managing three types of public resource: "virtual" resources defined by the software platform (such as software-based currency or reputation systems), physically constrained resources (such as limited RAM, CPU, bandwidth, and monthly server fees), and the consequences of antisocial behavior (such as vandalism, harassment, and hacking). These types span resource categories to include common pool resources, public goods, and pollution, all of which pose problems for collective action, the first two because they must be maximized, and the latter because it must be minimized [11]. These resources are very difficult to manage well. Vandalism and other forms of pollution are endemic to the platform, increasingly so in larger communities. And because the software is so resource intensive, a basic server can become strained with as few as 2 users, much less 200. Recommendations to server owners suggest that a server be provisioned with several Mbits of network bandwidth and 1 GB of RAM *per additional player* in order to provide users with a sufficiently responsive experience. In the face of CPU, RAM, and bandwidth limits, the only alternative to constraining user freedom is assuming the costs of operating a more powerful machine, and either providing it for free or looking to users for revenue streams such as donation drives, membership fees, paid advertisements, or premium services. On top of this, each community exists in a larger ecosystem of communities that are in competition for a fourth scarce resource, committed users, who are in demand because of the great potential value they can bring to the communities that manage to retain them. For example, in a remarkable investigation of competitive practices among professional video game servers, security researchers described hacker-for-hire schemes in which competing server administrators contract denial-of-service attacks designed specifically to frustrate and fragment their competition's user base [49]. These exploits are uncommon in the amateur-run (non-professional, not-for-profit, and generally smaller) servers we focus on in this work. Nevertheless, they successfully illustrate the sensitivity of server communities to resource challenges: an administrator who fails to provide necessary resources up to users' standards, whether physical or virtual, risks the desertion of their core group to a better-run server.

Fortunately, online communities are in a unique position to benefit from self-organized governance systems — formalized text policies and automated software rules. A study of Wikipedia's formal policy shows strong conservation of core rules, around which subsequent rule-making has organized itself [25]. In the communities we study, the challenges posed by resource-related social dilemmas have been made approachable with open-source catalogs of "plugins": modular programs that automatically implement rules and other political-economic constructs. These include temporary bans, full exile and blacklisting, hacking counter-measures, cheater detection, peer monitoring and reporting tools, surveillance tools, distribution of authority to trusted members, tools that reduce the number of game actions to those that are least computationally costly, and tools for keeping and restoring backups after failures and attacks. A plugin called Lockette instantiates private property to prevent stealing by implementing the idea of personal storage, one called WorldBorder conserves server disk resources by making virtual resources more limited, and another called LogBlock improves monitoring of all three types of resource by giving every possible virtual entity a public recorded history. Code in these communities is literally law [50]. And by "mixing and matching" many such rule systems, an administrator can implement virtually any governance system. In this work, we represent an administrator's governance style as a distribution over the types of governance plugin they have installed.

# Measures and predictions

In the communities we study, administrators start from a default state of a near absence of rules, and approach several types of resource challenge by selecting across several types of rule system: those that facilitate interpersonal communication, information diffusion, resource exchange, and top-down administrator control. All of these types have been implicated in improving outcomes in economic and collective action institutions [51-54].

Another key choice that administrators make in this platform is deciding the value of a mandatory server parameter: the maximum number of users who can be logged in simultaneously. We use this setting as a proxy for each administrator's desired target population size.

We define community success as the size of a server's "core group," the number of users who returned at least once a week for at least four consecutive weeks. In contrast to raw number of visits over a time period, return visits indicate a sustained level of interest and commitment (SI Text: Community Success). By measuring core group size with target size in mind, we define "success" with respect to administrator goals (Fig 1): a community with a core group of 4 (4 users who returned at least once a week for a month) is successful if its target size is 4, and not if its target size is 100.

The 5,216 independent communities in our dataset represent a wide range of sizes, with 1–30,000 confirmed visits per month, 1–3,100 unique visits, target size from 2–284 (median 6), and success (in terms of core group size) ranging from 0–400 (median 1). The median lifetime of these communities was 8 weeks.

With multiple communities of multiple orders of magnitude all overcoming similar problems in a competitive setting, this platform lets us test the relationship between institutional structure on governance success, as moderated by target population size. Following the claims of the theory of institutional diversity, that institutional complexity is a response to environment complexity [10,55], we measure how communities develop in terms of three dimensions of regime complexity: the number of rule systems they install (rule count), the heterogeneity of rule types they represent (rule diversity), and the purview of those rules, in terms of number of resource problem types (rule scope). Rule count gives a fairly literal representation of the size or extent of a server's policy apparatus, complexity in the sense of representing a large complex of rules. Rule diversity is the closest indication of institutional diversity: the variety of structures and rule types that an institution employs. For example, ideologically market-focused or authority-focused administrators might try to build servers that rely only on plugins that institute exchange mechanisms or that further empower the administrator. Their servers would show low institutional diversity, while a less discriminating or more pragmatic server with a variety of rule types would show high institutional diversity. Rule scope indicates the number of resource challenges that the administrator is explicitly attending to; are they focusing narrowly on just the problem of bad behavior, or server resources, or are they working in parallel to manage many? All three complexity measures capture facets of an institution's development or integration, with rule scope and diversity respectively capturing the complexity of the types of problem and solution, and rule count capturing the extent of the institution. As a fourth measure, we also calculate rule specialization, which indicates how unique a community's governance was in the population of communities, as a proxy for the role of niche size, a concept from organizational ecology [56].

Varying in size from a few members to thousands, in the governance choices that their administrators have made, and in the levels of success they have achieved, this population of amateur-run servers makes it possible to observe formal institutions in various stages of development. What governance features predict a community's success, and how do those features differ between large and small communities? Do larger or more successful communities have more developed or complex governance schemes? If the drivers of institutional development are historical accident and capture — an accumulation of vestiges and institutionalization of arbitrary power — than we shouldn't expect more complex governance regimes to be more successful, but we might expect complexity to increase with size. We would also expect complexity to increase in size under the competing theory, that complex, integrated regimes develop in response to the demands of a complex, interdependent environment. But even though an effect of size does not distinguish the theories, they do differ in their predictors of success. Under the institutional theories of the resource management perspective, institutions will be successful to the extent that they can respond effectively to the demands of their environment, and the environment will become more demanding as resource limits impose more onerous constraints on population size. Following both theories, we predict that all measures of complexity (rule count, rule diversity, and rule scope) will increase with increases in community population. Following the resource management literature, we also predict that these measures will increase with success, particularly among larger communities.

## Data and Methods

The online communities in our population are all servers of the multi-player "virtual world" video game Minecraft. Previous research with the game has focused on individual or group level game behavior, with a focus on creative play, collaboration, and engineering applications [57-61]. For our purposes, Minecraft stands out less for its qualities as a game *per se*, and more for the ecosystem of servers, tools, players, and practices that the player community has collectively built around it. By contrast to the business models supporting other games, where all servers are managed by a single professional entity, playing Minecraft with others usually means logging into an openly accessible server, somewhere in the world, that is being provided by an unpaid amateur without professional experience in governing strangers or managing server resources. Minecraft is an ideal domain for comparative institution-scale analysis because it is one of few games with a decentralized amateur-driven hosting model and a large user base. And it is ideal for testing questions of resource management and economic governance because administrators have autonomy, a clear goal, a wide variety of tools, and a challenging resource environment.

Independent of the game's specifics, merely logging in imposes a substantial burden on that server's computational resources, one that threatens to undermine the game experience for all. If the difficult nature of the bounded resources were not enough, the population also poses challenges. Most players are anonymous and often immature youth, two qualities that should make governance more challenging for a server administrator [62,63], and correspondingly more interesting for the study of successful resource management institutions.

Our analysis was based on a dataset of API queries from 370,000 Minecraft servers contacted between 2014/11 and 2016/11, several times daily. By default, these servers are publicly accessible via the Internet and do not have terms of use. Our scraper accessed each community for several public server performance statistics, including rules installed, maximum simultaneous users allowed (server "size"), and the anonymous IDs of users present. After filtering out disconnected servers (~220,000), those that did not survive for at least one month (~70,000), and those that did not report full governance information (~75,000), we had a corpus of 5,216 minimally viable, minimally comparable online server communities, 1,837 of which were also minimally successful (full detail in the SI Text: Data Processing). Part of minimum comparability is that we excluded large professional servers from our analysis, chiefly because their ultimate goal is not to build a community but to be profitable. This difference leads them to work to maximize impressions (unique rather than return visitors) and to focus on distinguishing themselves from other large servers, modifying the game environment and mechanics so heavily that they are scarcely recognizable as servers of Minecraft, in terms of the challenges they face or how they address them.

Administrators select software rules from a single central community-managed plugin repository. Within this system, each is assigned by its author to a category that describes what type of rule it is. We used these categories to classify rules into types, and to count each community's rules by its governance characteristics. A community's rule count is the sum of plugins over all three resource types. A community's rule diversity and rule scope (resource diversity) are the ecological variety (number of types) represented by its total system of rules: a server with no governance plugins has ecological variety of zero, while a server with at least one plugin in two different categories has variety two. A server's rule specialization was the median, over all plugins, of the number of other servers that plugin was observed on.

Our main analyses regress core group size and the $log_2$ of population maximum against these four measures of institutional diversity and the interactions of each with community target size, and several basic covariates (Table 2, SI Text: Data Analysis). To cancel the leverage that unsuccessful communities had on models of population size (which did not control for core group), we conducted all tests on population maximum on only the subset of 1800 minimally successful communities (core group size > 1).

## Results

Although our analysis is correlational, we present a causal reading of our results. Specifically, we interpret our results as estimating the effects of governance regime as a necessary condition for success. We support this interpretation on the grounds that domain constraints prevent a community from growing its core group without having overcome endemic resource problems. Plentiful resources are a prerequisite to the successful cultivation of a large core group. As large population sizes exacerbate resource management problems [48], communities are under more pressure to manage resources well. It is important to keep in mind the difficulty of collective action in this setting and others [32,64]. Within our own sample, only 35% of administrators ever recruit a core group larger than themselves, and fewer than 5% build a core group of more than four. With more users in-game resources are extracted at a more aggressive rate, server CPU, RAM, and bandwidth approach their physical limits, and the probability of malicious users increases. Any one of these may drive community members away. With the intense selection pressure on this population of communities, especially large ones, it is not feasible for servers to attain our definition of community building success without having already achieved success at managing all of the resource problems that constrain their growth.

We first consider how governance features are correlated with server population size. As this value is actually a desired maximum set by each administrator, predictors of server size tell us how administrators' understandings of effective governance style change with their intended community size, among communities that were at least minimally successful. As the governance features of interest are likely to covary, and we were unconstrained by theories for which features might affect which others, we ran several models, building up from models that test each feature individually (plus controls), to a full model including all features. Considered together, the single-variable models are consistent with the interpretation that the three complexity measures all correlate positively with maximum server size (tested separately; Fig 1; all $p<0.001$; Table 1 models 2–5) and that rule specialization does not ($p=0.43$). The full model (Table 1 model 6) complicates this picture only slightly. According to both the single and full models, rule count has a very robust positive correlation with server size (Table 1 models 2 and 6), suggesting that minimally successful servers consistently install more governance plugins as their target size increases. The full model also supports the lack of effect of a server's specialization, in terms of the uniqueness of its rules relative to those in use by other servers ($p=2.0$), and complicates the interpretations rule scope and rule diversity, the first of which is effectively insignificant considered in combination with the other features ($p=0.03$), and the second of which remains significant, but flips from positive to negative, suggesting that rule diversity actually decreases with server size ($p<0.001$). These results are likely due to covariance patterns between diversity and scope, which in turn may give hints as to the mechanisms behind the correlations of these features with maximum server size.

Given an overall increase of institutional complexity with intended server size, we also consider the effects of these same features on community success, as operationalized by the size of a server community's core group (again, the number of visitors who returned to the community at least once a week over a month; Fig 1 and Table 2). In order to capture both the overall effect on success, and any contingencies of these effects with size, our models fit both main effects and interactions with maximum server size. Socioecological theories, such as those from the resource management perspective, predict a positive relationship between governance complexity and community success in such a complex resource setting, one that we interpret directionally as a positive effect of complexity on success.

We find that several indicators of institutional complexity are statistically significant predictors of community success, but only in interaction with a server's maximum population size. In the models testing governance features individually, the main effects of rule count, diversity, scope, and specialization on success are above our $p<0.001$ threshold for statistical significance ($p=0.02$, $p=0.005$, $p=0.06$, $p=0.22$; Table 2 models 2–5). However, considering interactions with community size reveals significant positive effects of two features in interaction with size: rule count and rule scope (both $p<0.001$; models 2 and 5). Deploying a greater overall amount of formal (code-mediated) rules increases success among larger servers, as does deploying rules that address a greater variety of resource types. Considering the full model (Table 2 model 6), with all four governance features together, again supports the overall robustness of the positive effect of rule count in interaction with size ($p<0.001$), the clear insignificance of rule specialization ($p=0.87$), and a colinearity relation between rule diversity and scope that may result from the specific mechanisms connecting regime style to success. In the full model, rule scope becomes insignificant ($p=0.34$) and the effect of rule diversity on success with size becomes significantly negative ($p<0.001$).

We next look more closely into rule scope and rule diversity, to determine if certain rule types, or attention to certain types of resources, are significantly associated with size or success (Fig 2). Within rule types, we find that larger communities rely to a greater extent on rules that further empower a server's central administrator ($p<0.001$, Table 3 model 1), but no significant effects on community success, either alone or in interaction with size (Table 3 model 2). Investigating within resource types we find statistically significant positive correlations between size and rules for behavior and physical resources (both *p*<0.001, Table 4 model 1). Physical resource management rules also have a significant positive effect on success, but only in interaction with size, such that larger servers that focus on the governance of computational resources are significantly more successful (*p*<0.001, Table 4 model 2).

Under our interpretation, a community's complexity or specific style of governance has increasing influence on its likelihood of succeeding or failing as it aspires to be larger. Because larger populations exacerbate each of the major resource problems facing servers, large communities are more susceptible to spontaneous failures of collective action, and seem to require more intentional, complex, integrated governance in order to succeed at recruiting and maintaining a sizeable core group. Overall, our results are consistent with resource management theories that complex governance schemes emerge in response to the demands of complex resource environments.

## Discussion

What drives small, rudimentary informal institutions to develop themselves into integrated, formal governance systems? In the context of online communities, the increases we find in the number and scope of rules with success are consistent with institutional diversity's prediction that regime complexity emerges among successful servers as an adaptation to a complex environment. And our context-sensitive definition of "success," as relative to each community's own target size, implicates population size as a major driver of regime complexity.

Our findings also support the idea that communities benefit from a strong administrator. Although the ideas of community and small-scale governance often imply democratic aims, the communities we study here are not democracies. By default, a single administrator maintains complete control and, as we show, their power and authority increases with size. While it may be tempting to interpret this as an emergence of autocratic rule, it is important to be mindful of the evolutionary dynamic that binds these communities: communities actively compete for users, and users have unrestricted freedom to "vote with their feet" at negligible or relatively low cost (SI Text: Minecraft ecosystem) [65]. While their investments in a community may increase their apprehension at leaving, or their tolerance for bad administrators, the stakes of abandoning game constructs is ultimately small relative to the stakes that people face in more familiar applications of market theories of governance: the choice of what city to move to, what nation to immigrate to, or what representative to elect. Furthermore, players' IDs persist across servers, making it possible for them to retain certain types of value across servers, such as friendships and social capital. The persistence of player IDs makes it much easier for players to maintain relationships across servers.

Under the competitive conditions that servers endure, the best explanation for the value of concentrated authority is that users prefer it, and select into servers that exhibit it. Because we, like those who formulated the hypothesis of institutional diversity [7], are subject to romanticizing decentralization of political authority [66,67], this finding reminds us of the value of leadership in collective action settings, and it favors various economic theories of the state [15,68], including the utopian thesis of Nozick [4] that when tyrants must compete, market forces can drive them to govern as if they were benevolent. Therefore we propose that the relationship of increased administrator power to community success may not reflect an increase in, or preference for, authoritarianism.

## Limitations

The resource management framework we use, developed by the community of scholars around Elinor and Vincent Ostrom, were developed precisely so that insights from one socioecological setting, such as common pasture, could generalize to vastly different settings — fisheries, forests, irrigation systems, or online communities—each with their own peculiarities [32,69]. In all of these domains, and in Minecraft as well, environmental conditions create social dilemmas around valuable resources, and agents with high motivation and clear goals create institutions whose proper functioning aligns private goals with the public good. In these general terms, a multiplayer video game like Minecraft is as valid a source of generalizable insights as any other ecosystem. Still, there are bound to be properties of virtual communities that impede the generalizability of existing theory. For example, an alternate explanation for one finding, that most effort is devoted to managing bad behavior, may be less due to general features of institutions and more due to Minecraft's demographics, which recall William Golding's *Lord of the Flies*. And generally, as high-stakes as it is for a game server to fail, games are almost by definition lower-stakes than most other applications of institutional theory. Logging out of a server or losing a virtual good are generally less costly than emigrating from a nation or losing a livelihood. But as long as the stakes are high enough that resource scarcity poses a threat to users, tests of general theory in this domain remain justified.

Administrators' motivations to voluntarily incur the many costs of server management are affected by many unobservables that we do not investigate. Similarly, users are driven by many motives in how they choose which servers to explore and commit to, and may not experience the choice, for example, to leave an underperforming server as low or no cost. Still, we defend a rational choice framework for administrators, the agents analyzed in this work, because of the costs in time and effort they incur by choosing, for whatever reason, to opt into the difficult task of server management.

Our causal interpretation of our finding depends on the assumption that servers cannot maintain a core group if they are struggling with resource availability problems. And such problems are common: a small number of malicious or even naïve users can cripple server performance and drive core users away. Of course, in the absence of random assignment it is impossible to defend a causal argument definitively; it may be that having a large core group causes integrated governance effort, or that third variables like charismatic leadership cause both large core groups and integrated governance. These results are of no less interest if the causal direction we impose is invalid:

alternative interpretations are also of immense theoretic interest for the emergent, scalable collective action they imply.

# Conclusion

Online amateur institutions such as game servers, blockchains, wikis, forums, and social networks have become an exciting proving ground for researchers to scale up traditional comparative institutional analyses [43]. But, much more importantly, they are a proving ground for amateurs to develop their leadership abilities, and aptitude for the basic skills of democratic participation. This work illustrates the opportunities that general resource management frameworks and online sociotechnical systems promise each other: the former for foregrounding resource management as a constraint that can unify governance perspectives on online institutions, the latter as a source of population-of-population datasets that improve the quality, quantity, and pace of insights into the nature of effective institutions [35]. At a time when technology is empowering and connecting more people, institutional perspectives on online communities contribute to a unified view of human institutional development — trade treaties to town halls, businesses to bulletin boards, *al hima* to *harambee* — and help more people to benefit from the sciences of social design.

# Acknowledgments

The authors wish to acknowledge attendees of the Science of Counter Earth Workshop at Dartmouth College's Neukom Institute, Brian Keegan, Peter Krafft, Simon Marti, Michael Cox, Mubbasir Kapadia, Alexander Shoulson, Stephan Müller, Dirk Helbing, Markus Gross, the laboratory of Luke Chang, and the support of the Neukom Institute's Neukom Postdoctoral Fellowship. This work was conducted using publicly available data, which we make available.

# Figures

**Fig 1. Most communities are small and unsuccessful. Larger successful communities have more rules governing more kinds of resources.**

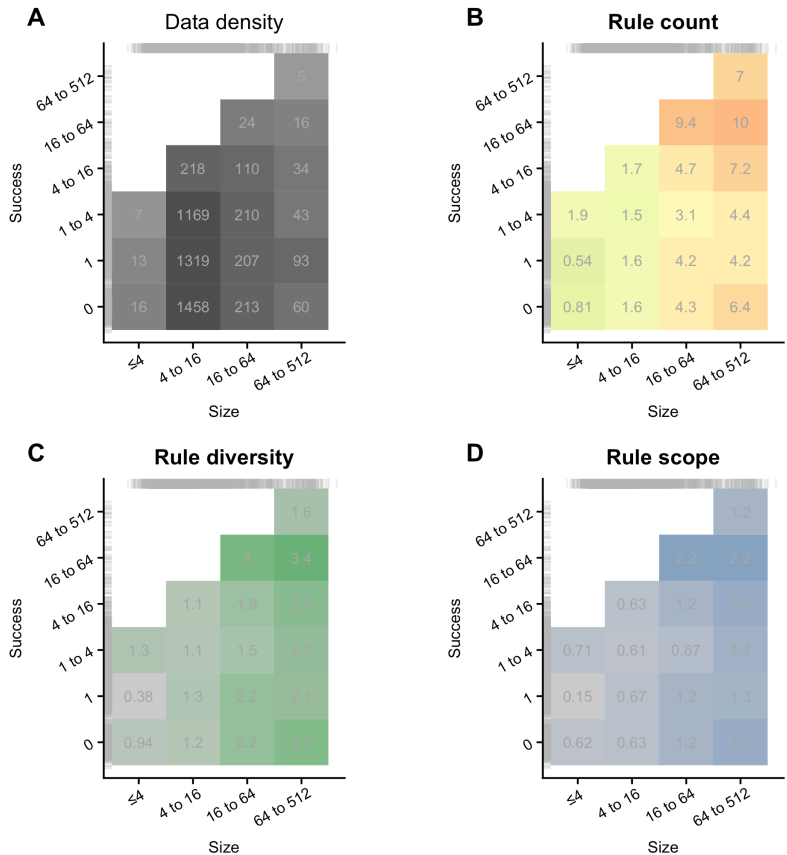

We analyze 5,200 amateur-run web server communities. Each server is operated by an administrator who makes all governance decisions. Among these decisions is the server *size (x-axis)*, the server's maximum number of users who may participate at any moment. This number represents an administrator's desired community size and puts a practical upper bound on the community's core group or *success*: the number of users who return to the community regularly (*y-axis; all plots*). Beyond return visits, unique monthly visits to many of these communities exceed the thousands. **A.** We summarize the data in a 2D histogram of all communities binned by success and size, with each bin reporting the number of communities within the given range, and marginals represented by grey ticks. Most communities have size 4–16, and most fail to grow a core group larger than one. The most interesting communities, those with the largest core group for their class, are along the diagonal upper edge of each plot. A bin's shade of grey, its number label, and the marginals all communicate the same distributional information redundantly: the count of communities by size and success. **B.** Administrators select their community's governance regime by installing combinations of software modules that implement rule systems. This panel shows the mean number of rules in use by communities in a bin. **C.** and **D.** All rules address some resource problem with some kind of rule. There are different problems and different rules (Fig 2), and we plot diversity metrics over them. Panel C shows that large successful communities use a greater variety of rules types ("rule diversity"). Panel D shows that they attend to a greater variety of resource problems ("rule scope").

**Fig 2. Larger successful communities use rule systems with more types of rules governing more types of resources. Actively managing physical server resources increases success with size.**

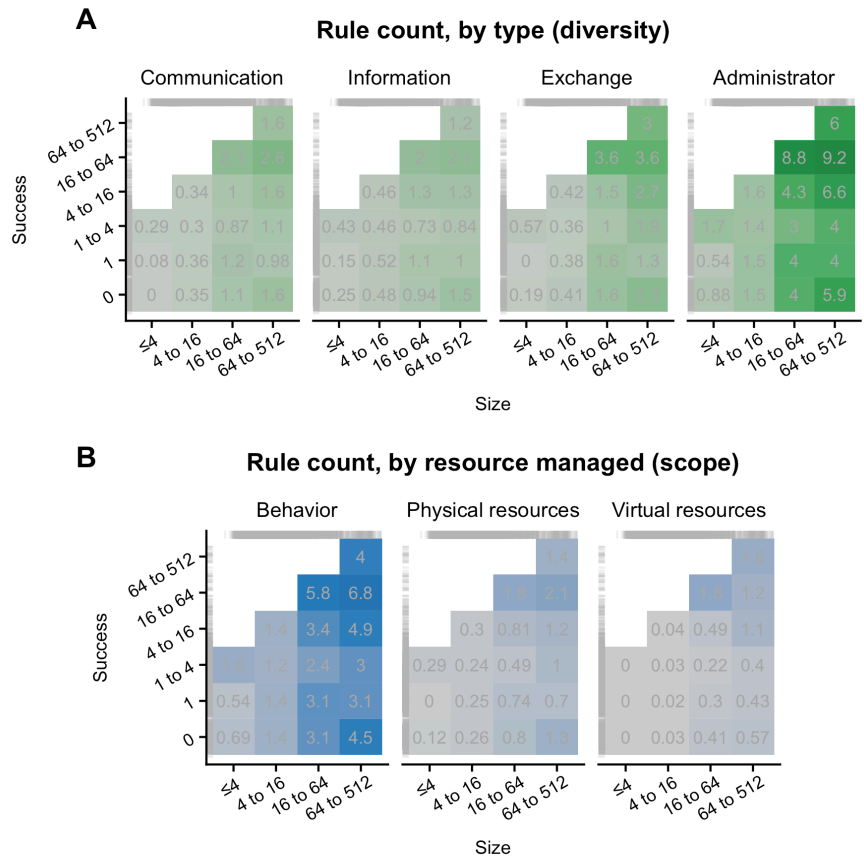

**Each plot shows the mean number of rules per bin, per rule type (A)** or target resource (**B**). The most common type of rule extends an administrator's power over their server. The resource challenge that attracts the greatest number of rules is the management of bad behavior. These two types, as well as rules that manage physical resources, increase in use significantly with population maximum (p<0.001). As their shared colors indicate, Fig 1C shows the diversity across the 4 plots of A, and Fig 1D represents the data over the 3 plots of B. For reference, both rows of figures roughly sum to Fig 1B.

# Tables

**Table 1. Models of population size. Rule count and other measures of regime complexity are greater in larger minimally successful communities.**

|  | Dependent variable | | | | | |
|---|---|---|---|---|---|---|
|  | Pop. max. | | | | | |
|  | (1) controls | (2) ctrls+1feat | (3) ctrls+1feat | (4) ctrls+1feat | (5) ctrls+1feat | (6) full |
| Intercept | 3.65*** (0.25) | 3.65*** (0.24) | 3.58*** (0.25) | 3.65*** (0.25) | 3.64*** (0.25) | 3.73*** (0.24) |
| API richness | -0.31*** (0.07) | -0.31*** (0.07) | -0.31*** (0.07) | -0.31*** (0.07) | -0.33*** (0.07) | -0.32*** (0.07) |
| Software count | 0.02*** (0.001) | 0.01*** (0.001) | 0.02*** (0.001) | 0.02*** (0.001) | 0.01*** (0.001) | 0.01*** (0.001) |
| Week | 1.49*** (0.15) | 1.24*** (0.15) | 1.35*** (0.15) | 1.49*** (0.15) | 1.29*** (0.15) | 1.22*** (0.15) |
| Weeks up | -0.02*** (0.002) | -0.02*** (0.002) | -0.02*** (0.002) | -0.02*** (0.002) | -0.02*** (0.002) | -0.02*** (0.002) |
| **Rule count** |  | **0.09*** (0.01)** |  |  |  | **0.11*** (0.02)** |
| **Rule diversity** |  |  | **0.13*** (0.02)** |  |  | **–0.14*** (0.04)** |
| Rule specialization |  |  |  | 0.63 (0.80) |  | 1.00 (0.78) |
| Rule scope |  |  |  |  | **0.28*** (0.03)** | 0.15* (0.07) |
| Observations | 1,837 | 1,837 | 1,837 | 1,837 | 1,837 | 1,837 |
| $R^2$ | 0.19 | 0.24 | 0.21 | 0.19 | 0.22 | 0.24 |
| Log Likelihood | –2,953 | –2,897 | –2,929 | –2,952 | –2,916 | -2,890 |

*p<0.05  **p<0.01  ***p<0.001

A server's max. population size is the maximum number of users who can be logged in simultaneously, and it proxies the administrator's desired community size. Columns report regressions fitting features of 1,837 minimally successful communities to $\log_2$ of population size. Predictors of interest are high-level features of the rule systems installed by communities. Model 1 fits only controls, models 2–5 fit each institutional feature individually, and model 6 fits all predictors. Control variables, represented in the tables in grey, include the richness of a community's voluntary public API reporting, its total number of installed plugins (both governance related and non-governance related), a date of the server's measured activity in weeks, and its duration to date, in weeks. A hairline rule separates theoretically relevant variables from controls.

**Table 2. Predictors of community success, in terms of core group size. Effect of rule count and other measures of regime complexity interacts with population size.**

|  | Dependent variable |  |  |  |  |  |
|---|---|---|---|---|---|---|
|  | Core group size |  |  |  |  |  |
|  | (1) controls | (2) ctrls+1feat | (3) ctrls+1feat | (4) ctrls+1feat | (5) ctrls+1feat | (6) full |
| Intercept | -0.97*** (0.15) | -0.73*** (0.15) | -0.90*** (0.16) | -0.97*** (0.15) | -0.85*** (0.16) | -0.84*** (0.15) |
| API richness | 0.01 (0.04) | -0.02 (0.04) | 0.005 (0.04) | 0.01 (0.04) | -0.01 (0.04) | 0.01 (0.04) |
| Software count | 0.01*** (0.001) | 0.01*** (0.001) | 0.01*** (0.001) | 0.01*** (0.001) | 0.01*** (0.001) | 0.01*** (0.001) |
| Week | 0.17 (0.09) | 0.18* (0.09) | 0.19* (0.09) | 0.17* (0.09) | 0.18* (0.09) | 0.17 (0.09) |
| Weeks up | 0.01*** (0.001) | 0.01*** (0.001) | 0.01*** (0.001) | 0.01*** (0.001) | 0.01*** (0.001) | 0.01*** (0.001) |
| Pop. max. | 0.27*** (0.01) | 0.25*** (0.01) | 0.27*** (0.01) | 0.27*** (0.01) | 0.26*** (0.01) | 0.26*** (0.01) |
| Rule count |  | -0.01* (0.01) |  |  |  | 0.0002 (0.01) |
| Pop. max. × Rule count |  | **0.02*** (0.003)** |  |  |  | **0.03*** (0.005)** |
| Rule diversity |  |  | -0.03** (0.01) |  |  | -0.06* (0.03) |
| Pop. max. × Rule diversity |  |  | 0.01* (0.01) |  |  | **-0.05*** (0.02)** |
| Rule specialization |  |  |  | -0.57 (0.47) |  | -0.48 (0.46) |
| Pop. max. × Specialization |  |  |  | -0.10 (0.16) |  | 0.02 (0.16) |
| Rule scope |  |  |  |  | -0.04 (0.02) | 0.05 (0.05) |
| Pop. max. × Rule scope |  |  |  |  | **0.04*** (0.01)** | -0.02 (0.03) |
| Observations | 5,216 | 5,216 | 5,216 | 5,216 | 5,216 | 5,216 |
| Log Likelihood | -8,494 | -8,472 | -8,489 | -8,492 | -8,488 | -8,449 |
| Deviance | 5,190 | 5,200 | 5,189 | 5,190 | 5,191 | 5,208 |

*p<0.05  **p<0.01  ***p<0.001

A server's core group size is the number of users who returned at least once a week for a month, and it quantifies success at community building. Columns report negative binomial regressions fitting features of 5,216 communities to core group size. Predictors of interest are high-level features of the rule systems installed by communities. Model 1 fits only controls, models 2–5 fit each institutional feature individually, and with its interaction with population size, and model 6 fits all predictors. A hairline rule separates theoretically relevant variables from controls.

**Table 3. Rule types as predictors of population size and core group size. Rules that empower administrators are more likely among large servers.**

|  | Dependent variable | |
|---|---|---|
|  | Pop. max. | Core group size |
|  | *(1) rule types* | *(2) rule types* |
| Intercept | 2.91*** (0.04) | -0.59*** (0.04) |
| Software count | 0.01*** (0.001) | 0.01*** (0.001) |
| Pop. max. |  | 0.25*** (0.01) |
| Rule$_{Communication}$ | 0.09 (0.05) | -0.04 (0.03) |
| Rule$_{Information}$ | -0.09 (0.05) | 0.03 (0.04) |
| Rule$_{Exchange}$ | 0.01 (0.05) | 0.01 (0.03) |
| Rule$_{Administrator}$ | **0.11*** (0.03)** | -0.02 (0.02) |
| Pop. max. × Rule$_{Communication}$ |  | 0.0002 (0.02) |
| Pop. max. × Rule$_{Information}$ |  | 0.01 (0.02) |
| Pop. max. × Rule$_{Exchange}$ |  | 0.03 (0.02) |
| Pop. max. × Rule$_{Administrator}$ |  | 0.001 (0.01) |
| Observations | 1,837 | 5,216 |
| $R^2$ | 0.21 |  |
| Log Likelihood | -2,936 | -8,540 |

*p<0.05  **p<0.01  ***p<0.001

A hairline rule separates theoretically relevant variables from controls. Note that maximum population size is the dependent variable in the first model and an independent in the second.

**Table 4. The resources targeted by rules, as predictors of population size and core group size. Rules that manage a server's computational resources are increasingly successful with size.**

|  | Dependent variable | |
|---|---|---|
|  | Pop. max. | Core group size |
|  | (15) resource types | (16) resource types |
| Intercept | 2.93*** (0.04) | -0.55*** (0.04) |
| Software count | 0.01*** (0.001) | 0.01*** (0.001) |
| Pop. max. |  | 0.24*** (0.01) |
| Resource$_{Grief}$ | **0.08*** (0.01)** | -0.03** (0.01) |
| Resource$_{RealWorld}$ | **0.42*** (0.05)** | 0.03 (0.06) |
| Resource$_{InGame}$ | 0.03 (0.04) | 0.01 (0.02) |
| Pop. max. × Resource$_{Grief}$ |  | 0.004 (0.01) |
| Pop. max. × Resource$_{RealWorld}$ |  | **0.08*** (0.02)** |
| Pop. max. × Resource$_{InGame}$ |  | -0.01 (0.01) |
| Observations | 1,837 | 5,216 |
| $R^2$ | 0.22 |  |
| Log Likelihood | -2,924 | -8,521 |

*p<0.05  **p<0.01  ***p<0.001

A hairline rule separates theoretically relevant variables from controls. Note that maximum population size is the dependent variable in the first model and an independent in the second.

# Supplement for

# Emergence of integrated institutions in a large population of self-governing communities

Seth Frey[*,1,2], Robert W Sumner[3]

[1] Communication Department, University of California Davis, Davis, CA, USA

[2] Neukom Institute for Computational Science, Dartmouth College, Hanover, NH, USA

[3] Computer Science, ETH Zurich, Zürich, Switzerland

# Contents

# Supplementary text

## Data context

### Self-hosting on the Internet

A server is a computer that provides services that other computers can access through the physical network structure of the Internet. These services can be of many types, and while the most familiar type of service is of web pages to the World Wide Web, other services support other modes of interaction. For example, multi-player game servers can foster social interactions in virtual worlds. Even though most of our interactions on the Internet are through professionally managed servers, the barriers to amateur servers are low, and it is increasingly common for public online community services to be hosted by non-professionals with only a minimum of skill at server management. Amateur hosting brings the time and money costs of maintenance to a server's main administrator, likely an individual with access to limited resources. Any efforts that that administrator makes to solicit resource provisioning assistance from users makes their behavior relevant to questions of governance posed by the frameworks of self-organizing resource management.

### Minecraft

Minecraft is a multi-player virtual world game well-known for its open unplotted structure and immense popularity: it is the second best selling of all time behind Tetris. It stands out from other games as a domain for institutional research, not for anything about the internal features of the game as a game, but for the culture around it and technical features of how it is made available to its players. Unlike most other multiplayer games, which are hosted on central severs controlled by the game developer, Minecraft permits a decentralized hosting model in which amateurs host instantiations of the game on servers they manage.

This study is based on data from a continuous survey of data from 300,000 Minecraft server IPs over two years. Regarding privacy, we only accessed data that servers made public. By default, servers provide basic information about their current state through an API that is publicly accessible via the Internet. Also by default, they have no terms of use governing their APIs, although some deviations from this default are likely, for those servers that may have published terms of their own. It was not feasible to determine which might have published terms, much less whether any terms had provisions that were relevant to how we scraped. However, we only accessed that data that was provided publicly by each server's API, subject to whatever constraints that server might have imposed on our queries; if a server did not provide a datum through its API, we did not record it.

**Minecraft players**

From what studies exist, scientific and otherwise, we can build a coarse sense of the demographics of the Minecraft user base. We estimate that approximately 80–90% of players are male and that the median player is a young adult (age 19) [70,71].[1] Out of the global community of users, the majority are based in the US and northern Europe. We do not have demographic statistics on the subset of players who are also administrators, but because of the technical skills and monetary costs required to administer a server, administrators are likely slightly older than other players.

Another dimension of social context to consider is that users are often acquainted in real life. One study of smaller game teams found that 75% of virtual colleagues are real-world friends or family [72]. By reducing anonymity and freedom from consequences for antisocial behavior, real world acquaintance raises the accountability of users to each other, and may to some extent supplant the function of installed rules. We predict that that non-observable informal avenue for norm enforcement is more likely in smaller communities, which provides an alternate, but non-competing hypothesis for our finding that software rules increase in number with population maximum.

**Minecraft servers**

To host a Minecraft community, a prospective administrator rents or buys access to a server, installs and configures the game's server software, installs plugins, and may then log in. When the server is started, a unique world is generated from the "seed" of a single number, usually randomly generated.

---

[1] for the results of an informal community solicited survey conducted in 2011, see http://i.imgur.com/xmL5Y.png

Administrators create rules and other modifications to their server by installing plugins. Plugins are small programs that are designed to modify server functionality in a well-defined way. Administrators may certainly deliberate with their users before installing new plugins, but they are ultimately free to act unilaterally, and typically do. While many plugins add frivolous functionality (in the literal sense that they foster play), many are explicitly developed to aid in the governance of specific resources, either by implementing single rules (like narrow proscriptions on possible behaviors) or larger self-contained institutions (such as markets or social hierarchies). While play is an interesting subject, the focus on this work is on the resource-related social dilemmas that administrators overcome in the process of supporting play and other user interactions.

Administrators must then attract a user base. They may recruit from among personally known peers, and they often advertise their community at a larger scale on web sites that provide users with searchable lists of servers, explained below.

**Minecraft ecosystem**

*Server lists and the administrator community.* Communities compete for users through "server list" sites, which players use to search for communities that match their interests. These sites work by aggregating URLs and other data from hundreds of thousands of active servers.[2] To improve the richness of this system, developers devised a public query API that game servers use to expose basic realtime statistics about the state and health of their community. This information includes a server's version number, its currently active population (with unique public user IDs), its maximum population, and so on. Server list sites are constantly querying game servers for this information in order to provide quality signals about each community they list. This query API information is made available by default on all servers, and it is the basis of our information about the governance style and success of each server in the corpus.

---

[2] such as http://minecraftservers.org

Because servers compete for users, and because users can join and leave voluntarily at low or no cost, the Minecraft ecosystem unwittingly implements the "market for tyrants" theory of utopia described by Robert Nozick in *Anarchy, The State, and Utopia* [4]. We use this characteristic to explain why observed governance characteristics in our population may reflect the institutional preferences of users rather than those of administrators. Others have shown that even leaders with a history of corruption can be incentivized to work in the interests of those they serve [73].

*Plugin sites and the developer community.* Use of server lists, and competition between servers, are dimensions of the larger ecosystem of code, culture, and communities within which each server operates. Another dimension important to this study are "mod" sites[3] where developers have converged to create a central repository of open source plugins.

In games such as Minecraft, collections of plugins grow into fully fledged ecosystems as a platform's power users coordinate in the programming and free distribution of useful tools. The broader Minecraft community had developed almost 20,000 plugins by 2016, although the top 500 accounted for nearly 90% of plugins in use.

To help administrators make sense of the large number of modifications, developers are required to assign each plugin they write to at least one pre-specified category. We use these categories as the basis of rules types, described in full detail in the Data Constructs section.

# Data constructs
## Administrators

One fundamental assumption of our analysis is that administrators are motivated to build community and overcome the social dilemmas that attend community building and resource sharing. We assume that they pursue this goal by means they consider effective. Server administration is costly in time and money, and the game can be played in a "single-player" mode without a server, so we infer that administrators go to the additional difficulty of enabling social play over the Internet because it is worth the costs.

---
[3] such as https://mods.curse.com/bukkit-plugins/minecraft/

By contrast to administrators, normal players' motivations are more opaque and varied. For this reason, we structured the analysis so that our conclusions require boundedly rational choice from administrators, not users.

A less vital assumption is that each server has a single administrator, and that that administrator monopolizes administrative decision-making. While this structure is customary, it is not mandatory. It may be that administrators provide administrative access to multiple players, or that they consult their users before instituting changes. We cannot observe this, but because a server can ultimately only be run one way, the outcome chosen by a group of decision makers is still a comparable to that chosen by a single decision maker, and it is safe to model that group as if it is a unitary agent.

**Size**

There are many ways to define the size of a community, such as its number of monthly visits, number of unique visits, or return visits. Instead, we use the population maximum, the value of a server parameter that limits the maximum number of simultaneous users. A downside of the population maximum parameter is that it underestimates other more intuitive definitions of size. For example, a server with a maximum of 10 simultaneous may still be visited by thousands of unique players in a month. However, this measure also has advantages:

— Unlike other potential measures, which are based partly on the results of an administrator's behavior over time, population maximum is set directly by the administrator upon installing the server. It captures an administrator's intentions for their server.
— Servers can handle indefinite load when that load is distributed sparsely over time. A server starts to encounter its performance limits only as it struggles with simultaneous requests. These performance limits are due to finite CPU, RAM, and bandwidth (some of the resources that administrators manage). The population maximum is a satisfactory definition of server size in part because it is directly subject to resource constraints and must be set with them in mind.

Administrators must balance their desired community size against their server's performance limits, and an administrator's choice to aspire toward a large community comes with the knowledge that they must arrange for sufficient resources to give that number of simultaneous users a good experience. The consequences of under-provisioning a high traffic server are network lag, low frame rate, unsynchronized interactions, lost connections, and other features that are known to quickly turn users away. Lag in particular, often referred to as "lagg" within video game communities, is a notorious turn-off, and is a major target of malicious agents seeking to disrupt a server.

— Because it must be manually updated, a server's population maximum is likely to remain stable and unchanged over months. When this important parameter does change, it is because the administrator intentionally changed it.

— Although it is very different in definition from the core group variable, population maximum also puts a soft upper bound on core group size. Out of 5,216 servers, we observe only 2 instances in which a community's core is larger than its maximum number of simultaneous users.

— Administrators do have incentives to set it honestly. If they set it lower than they desire, it will not be possible to host as many users as they desire. If they set it higher than they desire, they risk a sudden influx of users crashing their server or degrading the quality of game service below acceptable levels.

**Community success**

We evaluate servers in terms of the success of their administrators in recruiting a core group of committed community members. We define success specifically as the number of users who returned to that community at least once a week for a month. In contrast to raw number of visits over a time period, intermittent returns indicate a sustained level of interest and commitment in an environment in which many other communities are competing for each user's attention. Months are meaningful time units because they define the customary billing cycle for server hosting, because it takes about a month to bring a community to a mature state of development by the game's constructs, and because most servers endure for only 0–3 months.

With this measure of success, the basic unit of analysis is the server-month. An alternate quantification of community might have focused on the number of people who visited or returned over the entire lifetime of that community. However, because communities varied widely in their lifetimes, focusing on server-months improved the comparability of the longest to the shortest lived communities.

To make success relative to administrators' different goals, our figures consider each server's core group size plotted against its maximum population, and our models capture other interactions between these variables.

**Unit of analysis**

The servers in our sample varied widely in the number of weeks that they persisted. In order to control for the relative over-representation of longer lived servers (because they endured for more server months), we aggregated the data down to the scale of server-months and completed our dataset by selecting the most successful month of core group activity in the lifetime of each server. By selecting the month that a server reached its peak core group we improved the comparability of administrators by comparing them at their peak community building performance. This is the month in each server's lifetime when it is safest to assume that the administrator is highly motivated to accomplish the goal of building community. The cost of this choice is that this work fails to take advantage of the longitudinal nature of our raw dataset, a task we leave to future work. Alternative approaches would have been to select a random month, the very first month, or the very last month, or to control another way for the differences in sampling density of each server/administrator (and correlation of sampling density with success via longevity).

**Resource types**

Understanding a community's resources in these abstract terms makes community success amenable to frameworks for analyzing real world community resource management institutions such as local fisheries, irrigation systems, and forest management communities. In the case of Minecraft communities, administrators must work to manage public goods and common pool resources of three types: physical, virtual, and, more abstractly, antisocial behavior, a type of pollution.

*Physical resources.* The administrator usually has exclusive or de facto exclusive access to their server's overall status, underlying settings, and total available resources. Resource intensive applications such as Minecraft require increasing amounts of CPU, RAM, disk, and network bandwidth for each additional user on the server. Because an administrator, by default, cannot keep a player from logging on, a server's computing resources can be classed with common pool resources or congestable public goods. And because Minecraft defines a large, dynamic, real-time, 3D virtual world, these resources are very easily stressed, with standard recommendations prescribing about 5 GB of disk space, 1–3 Mbits/s of up- and download bandwidth, and 1 GB of RAM *per player* in order to provide players an acceptable level of moment-to-moment responsiveness.[4] These sometime prohibitive requirements can themselves quickly become inadequate in the presence of malicious or even naïve users who intentionally undermine service by triggering resource-intensive game events.

When computational resources are inadequate, the server will become unresponsive to player actions. This is called lag, and players are sensitive enough to it that it need only persist for a few minutes for them to leave for another server.

Additionally, administrators require monthly fees to maintain their connection to the Internet, whether hosted privately or through a firm. While it is possible to pay nothing to provide a public server, the typical amateur will begin paying about $10USD/month, and an administrator of a large, popular community may pay hundreds per month. While these figures may not sound like much compared to the millions invested in fisheries, for example, it is worth remembering that administrators are usually volunteers, and often teen-aged, and this study focuses on non-professional servers. Administrators are responsible for the payment of these fees by default, at which point the server itself is being provided as a public good to its players. When administrators attempt to raise fees from players, they are engaging with the well known problem of provisioning a public good. Players are incentivized to free-ride (it is, indeed, the norm), but failing to provide

---

[4] see, for example, https://minecraft.gamepedia.com/Server/Requirements/Dedicated

computing resources at sufficient quality or capacity undermines players' experiences and makes them likely to leave a community.

*Virtual resources.* Minecraft itself defines many virtual resources. We focus on in-game common pool resources that require both provisioning effort and responsible extraction. These are chests, crops, trees, and mobs. All of these are by default non-excludable, subtractable, and take effort to provide. Many plugins assist in the management of these resources, some by creating the idea of private property rights, others by improving logging of game events.

*Bad behavior.* Bad behavior is common enough in Minecraft that it is referred to within the community as "grief." Most (but not all) servers admonish players to not grief each other. Because the game's world is so open-ended, users can find many creative and hard-to-detect ways to harass each other. Harassment can take the form of chat-based insults or virtual violence, hacking a server, cheating by covertly installing performance aids on the client side, vandalism of the work of others, cursing, or otherwise behaving outside the bounds set by the administrator. Bad behavior can also affect server performance: malicious players will commonly start large virtual fires, grow a large virtual animal populations. These dynamic game entities require computational resources to represent individually, so if an attacker can arrange them to grow exponentially, even for a brief period, they can effectively deny service to the rest of the community.

The plugins of the Minecraft ecosystem provide a very wide range of partial solutions to the problems of bad behavior. These include temporary bans, full exile and blacklisting, hacking counter measures, cheater detection, peer monitoring and reporting tools, admin surveillance tools, distribution of authority to trusted members, and tools for keeping and restoring backups after attacks.

*No resource.* Many plugins in use by servers did not identify a resource, or did not identify themselves as relevant to governance. We excluded these from the analysis, except in the form of a control variable that tracked the raw number of plugins separately from the number of resource related plugins.

**Rule types**

We describe four types of software rule system recognized by the plugin developer community for addressing these various resource problems: "chat," "informational," "economy," and "admin tools." These strategies are not only common in Minecraft, but are recognized in the study of resource management generally.

*Chat plugins* are those that facilitate inter-player communication, either by expanding built-in capabilities, creating new communication channels, or improving the interoperability of multiple communication schemes. By default, users can only communicate via text chat, but various plugins enable the higher bandwidth of voice chat. Rules that improve communication have a recognized role in frameworks for institutional analysis [8], and unconstrained communication is known to be valuable in many resource governance settings [51], including Minecraft, which has plugins to support VOIP platforms such as Skype, Discord, and Teamspeak. In one case we observed, a server with a 16+ age limit verified compliance with that limit by requiring all users to communicate via voice chat.

*Informational plugins* are those that aggregate data for users and administrators and improve their knowledge of strategic decisions. Information rules are another type of rule that are recognized as important in frameworks for institutional analysis [8]. In Minecraft, the informational plugin Dynmap[5] help users monitor each other's locations and activities.

*Economy plugins* install private property rights and exchange mechanisms to permit players to trade in both virtual and physical resources. These plugins function as self-contained institution modules that administrators can use to ensure that certain resource distribution problems solve themselves. While most economy plugins restrict themselves to the distribution of virtual resources, some economy plugins, like Buycraft[6], help administrators cover server fees by selling social status or game objects.

---

[5] http://mods.curse.com/bukkit-plugins/minecraft/dynmap

*Administrative tools* are plugins that increase beyond the default the actions that the administrator can perform. These plugins help administrators exercise greater control over server state and player behavior. For example, the popular OpenInv[7] plugin allows administrators to covertly search and remove (potentially forbidden) belongings from their users' inventories.

It is important to note that not all governance need happen through plugins. If the members of a server's core group maintain real world relationships, they may have access to the much more compelling range of enforcement tools made available by physical co-presence. This is especially likely in the smallest server communities (of four and fewer), which are more likely to be managed by pre-existing real-world peers who could in principle resort to physical enforcement of their server's virtual rules. This caveat imposes some nuance on how we compare success in large and small communities, but to the extent that non-virtual enforcement characterizes small-community success, our key results are more rather than less likely to hold. In particular, this would not change our results for large servers, nor our observation that success looks different in small versus large servers.

## Data processing

We compiled a list of 376,576 Minecraft server addresses by querying for IPs from multiple sources[8] over two years. After filtering for communities that were minimally viable and minimally comparable, our final dataset was 5,216. We also identify, within these, the 1,837 minimally successful communities.

---

[6] http://mods.curse.com/bukkit-plugins/minecraft/buycraft

[7] http://mods.curse.com/bukkit-plugins/minecraft/openinv

[8] http://minecraftservers.org
  https://www.reddit.com/r/mcservers/ and
  https://www.shodan.io/search?query=+port%3A25565

**Minimally viable**

Of 376,000 servers, only 157,747 ever responded to our queries, and only 91,271 remained live for at least the span of one minimally costly month.

**Minimally comparable**

Our comparative analysis of governance schemes across communities required them to share a minimum number of common characteristics. All communities had to have a valid API, keep their functioning close to default, and report full community and governance data. This step reduced our sample from 91,271 to 5,216 server communities.

*API is trustworthy.* Servers' APIs are used to communicate information that prospective users use when deciding whether to join. But servers are customizable enough that administrators can undermine the ability of the platform's API to correctly report its performance statistics. We filtered out servers that were using special plugins that falsify this data in order to present more appealing statistics to players and server lists. We also filtered out servers reporting impossible values, such as a maximum population of zero or less.

"*Vanilla*". The world of Minecraft, being virtual, is virtually unlimited in its customizability. Indeed, with very simple modifications, a developer can rewrite physics and define away the finiteness of most of the finite resources at the core of this study. Finite or non-excludable resources can be made infinite or excludable, and therefore no longer susceptible to the collective action problems we study.

Deviations from vanilla play undermine key premises of this work, not to mention the comparability of servers to each other, and the reliability of their API statistics. Fortunately, and surprisingly, many administrators, especially those running small-scale amateur servers, choose to keep the game close to its default state, with resources that are indeed limited and in need of active management. They do so despite the "god-like" power that they have to define resource problems away entirely. Within the game community, servers whose mechanics remain close to default are called "vanilla" or "semi-vanilla" servers.

It is relatively straightforward to determine how far a server has drifted from vanilla play, as the most important deviations are implemented by easily identified plugins that we incorporate into our exclusion criteria. For example, some administrators install plugins that link many worlds into larger collections of servers. In these "multiverse" servers, API statistics fail to capture the extent of an administrator's community, so we also excluded them from our analyses.

*Full reporting.* The basic API reports statistics such as a server's maximum simultaneous users and list of users present. Beyond these basic statistics, administrators can opt in to a more full version of the API in order to more effectively compete for users. It is this non-default version of the API that reports a server's installed plugins, which we required in order to characterize a server's governance regime. Only about 10% of servers use the full version of the API. Because enabling the extended API improves the transparency of a community to prospective users, this filtering step may amplify the bias in our analysis for administrators who are motivated to recruit widely for visitors and core members. This source of bias would seems more likely to support than undermine our conclusions, particularly because it provides an additional filter for administrators that are highly motivated to overcome resource challenges toward building a devoted community.

## Minimally successful

An additional criterion that we included for a subset of our analyses is that a server was "minimally successful," with a core group of size greater than one. We used this subset of 1,837 out of 5,216 servers to test governance factors against group size. We did this because we did not want unsuccessful servers to drive the results of our tests, and we wanted to be able to interpret results over size as valid across the range a minimally successful servers. We did not include success as a covariate because our tests of success were already using population maximum as a covariate, and because administrators have more control over the value of the population maximum parameter than over then number of users that become community members. To manage the issues surrounding multiple tests of interacting constructs, we followed the procedure for multiple testing laid out by mediation analysis. Starting with models of population maximum, we moved to models of core group size that included population maximum as a control.

## Data analysis

Our central statistical tests regress core group size against four measures of regime complexity and the interactions of each with server target size, and several basic covariates. After describing the variables we describe the tests.

## Variables

DVs
— *Core group size.* This is the number of people who visited the server at least once a week during a server month. This is a main dependent of interest.
— *Population maximum.* This is the maximum number of users allowed simultaneously on a server. It is set by the administrator. Because some users set this astronomically high (enough to throw off our analysis), presumably to signal that they did not intend an upper limit, we capped this value at that maximum number of users ever observed simultaneously on any server. We $log_2$ transform the variable in our models (precisely *$log_2(x+1)$*, to theoretically allow for maximum population limits of size zero).

IVs: Controls
— *Week.* To capture large-scale fluctuations in the popularity of Minecraft, we included as a covariate the number of weeks from the Unix epoch, scaled.
— *Weeks up.* To capture potential effects of a server's age on its governance characteristics, we included a term for the number of weeks between the server month of interest and the date that our scraper first contacted the server.
— *API richness.* Servers vary in how much information they make available through various APIs, beyond the minimum required for our study. We included it in our models because it may be related to the willingness of users to visit a server, or to the level of involvement of the administrator.
— *Software count.* This is the number of plugins installed on the server, whether or not they are related to governance. This is in contrast to the rule count variable below, which counts only governance related plugins.
— *Population maximum.* Described above as a dependent. In addition to being a dependent of interest in itself, we also use this variable as a covariate and interaction variable in models of the other dependent, core group size. Because it effectively puts an upper bound on the size of a community's core group, the two are likely to be correlated.

IVs: Governance variables

— *Rule count.* This is the number of rules that are involved in the governance of some type of resource. It equals the sum of rules by resource type. It does not equal the sum of rules by rule type, because some rules are categorized to have multiple types, and some have none. Of course, some plugins install many rules, and some are very simple, but rule count nevertheless provides a proxy for the size of a server's governance infrastructure.
— *Rule diversity.* We identify plugins in terms of four rule types. Some servers use many rules of many types, while others use a smaller variety of types. We capture diversity of types in a variable by counting the number of types of rules represented on a server. This number of types is known in ecology as variety. We use variety instead of entropy because many servers have a very small number of rules, and entropy is known to be sensitive to small sample sizes and bins of size zero.
— *Rule scope (or "resource diversity").* Rule scope is the range of resource problems that a server is using plugins to address. We use the ecological variety, number of types, for the same reason that we use it to quantify rule diversity.
— *Rule specialization.* Because plugins are made publicly available, many communities use many of the same plugins, with the most popular plugins in use on tens of thousands of servers. The rule specialization of a server was the median of the number of other servers also using each plugin it was using. In our models we actually used the inverse of this count in order to define servers without any plugins as having rule specialization equal to zero. This fourth variable is insignificant in all tests. Exploratory analysis shows that its effects are accounted for by the software count control in single variable tests, and likely also by rule count in the full models.

**Model Specifications**

We used a mix of univariate and multivariate models on two dependent variables to understand the relationship between size, success, and various governance variables of interest. For all of these the unit of analysis was the server-month, with the dependent variables being the size and success of a server (population maximum and core group size). Though we do not characterize our analysis in terms of the statistical concepts of mediation or moderation, we followed the procedure established by moderated mediation analysis for testing over multiple dependent variables, first testing effects on population maximum, then testing effect on core group size with population maximum as a control variable and in interaction terms. Tables 1, 2, and 3 describe our models and their effects.

We modeled the logarithm of population maximum in five regressions, one that fit Week, Weeks up, API richness, and Software count, as well as Rule count, Rule diversity, Rule scope, and

Rule specialization, and four "single variable" models that fit the controls with just one of each of the governance variables. We used this range of tests to help us understand the relationship of the governance variables with each other. For example, rule scope is significant when modeled alone, but its effects seem to be accounted for by a combination of the other three (Table 1). And the positive coefficients for rule diversity alone become negative when considered in the context of the other variables, suggesting an overall increase in the diversity in rules with size, caused by the other variables, that is strong enough to mask a net decrease in diversity when they are all taken into account (Table 1). All models of population maximum were conducted on the subset of minimally successful communities — those with core groups > 1. This eliminates the inordinate influence of failed communities on our tests, and permits us to interpret any effects of size in terms of successful communities only.

We then modeled core group size with the same terms as above, plus the logarithm-plus-one of population maximum and its interactions with each of the four governance variables. Means before centering of each of the interacted variables were as follows: *Population maximum=3.17, Rule count=2.17, Rule diversity=1.36, Rule scope=0.76, Rule specialization=0.0059*. As above, we fit one full model with all four governance variables and four more models fitting each governance variable alone. Because core group size is a count of events — arrivals on a server — and because the baseline rate of these events varies by server, we used a negative binomial regression, in which parameter theta is understood as a dispersion over the lambdas of a population of Poisson processes.

For both the models of population maximum and core group size, Chi-squared tests of the full models against controls-only models showed a significant difference in variance explained, while tests of the single-variable models against the control models varied in their ability to exceed the $p<0.001$ threshold.

To better understand the rule count, rule diversity, and rule scope effects, we ran four more models (Tables 3–4),

- One linear regression fitting the four types of rule, plus a software count control, against population maximum.
- One negative binomial regression fitting the above variables plus population maximum and its interactions with the four types against core group size.
- One linear regression fitting the three types of resource, plus a software count control, against population maximum.
- One negative binomial regression fitting the above variables plus population maximum and its interactions with the three types against core group size.

**Plots**

The plots show the same data as fit to the models, except that we do not display "singular" bins containing only one server. This visualization choice led us to exclude two "above diagonal" outlier communities that were occupying two above diagonal bins. Including these does not influence the apparent conclusion of the plots, our analyses, or the manuscript. We excluded them to facilitate quick visual inspection of the complex 2D histogram plots that dominate our figures.

# Data availability

The data are available at https://dash.ucdavis.edu/stash/dataset/doi:10.25338/B8Q88S